# The Potential of Large Language Models in Supply Chain Management: Advancing Decision-Making, Efficiency, and Innovation

Raha Aghaei[1,2], Ali A. Kiaei[2,3]*, Mahnaz Boush[4]*, Javad Vahidi[1], Zeynab Barzegar[2], Mahan Rofoosheh[5]

Highlights:

- **Improved decision-making:** LLMs improve predictive analytics and operational efficiency of SCM.
- **Technological advances:** The introduction of transformer models such as BERT and GPT-3 is revolutionizing natural language processing in SCM.
- **Improved demand forecasting:** LLMs integrate different data sets for accurate, real-time demand forecasting and streamlined inventory management.
- **Optimized supplier relationships:** Automated communication and data-driven insights improve supplier performance analysis and risk management.
- **Efficient logistics:** real-time traffic and weather analyses enable dynamic route optimization and better fleet utilization.
- **Strategic integration:** By combining LLMs with IoT, blockchain, and robotics, intelligent and autonomous supply chains are created.

[1] School of Mathematics and Computer Science, Iran University of Science & Technology, Tehran, Iran
[2] Department of Computer engineering, Sharif University of Technology, Tehran, Iran
[3] (Correspondence: ali.kiaei@sharif.edu) Department of Artificial Intelligence in Medicine, Faculty of Advanced Technologies in Medicine, Iran University of Medical Sciences, Tehran, Iran
[4] (Correspondence: m.boush@sbmu.ac.ir) Cellular and Molecular Biology Research Center, Shahid Beheshti University of Medical Sciences, Tehran, Iran
[5] Computer engineering group, Alborz Vocational Technical University, Alborz, Iran

# Abstract

The integration of large language models (LLMs) into supply chain management (SCM) is revolutionizing the industry by improving decision-making, predictive analytics, and operational efficiency. This white paper explores the transformative impact of LLMs on various SCM functions, including demand forecasting, inventory management, supplier relationship management, and logistics optimization. By leveraging advanced data analytics and real-time insights, LLMs enable organizations to optimize resources, reduce costs, and improve responsiveness to market changes.

Key findings highlight the benefits of integrating LLMs with emerging technologies such as IoT, blockchain, and robotics, which together create smarter and more autonomous supply chains. Ethical considerations, including bias mitigation and data protection, are taken into account to ensure fair and transparent AI practices. In addition, the paper discusses the need to educate the workforce on how to manage new AI-driven processes and the long-term strategic benefits of adopting LLMs.

Strategic recommendations for SCM professionals include investing in high-quality data management, promoting cross-functional collaboration, and aligning LLM initiatives with overall business goals. The findings highlight the potential of LLMs to drive innovation, sustainability, and competitive advantage in the ever-changing supply chain management landscape.

# 1- Introduction to the Large Language Models in SCM

The dynamic and complex nature of supply chain management (SCM) requires advanced tools and technologies to increase efficiency, reduce costs, and improve decision-making. Large Language Models (LLMs), a breakthrough in the field of artificial intelligence, have proven to be powerful advantages in achieving these goals. These models, which excel at natural language processing and understanding, have found applications in various aspects of SCM, from demand forecasting and inventory management to logistics optimization and supplier relationship management.

In pure AI, there are recent papers which stretch to all directions and prefect the passing of LLMs into various application fields such as IoT, or general practice., Neither is natural. [1], [2], [3], [4], [5], [6], [7], [8], [9], [10], [10], [11]

On the other hand, LLMs are starting to become widespread in various new fields. RAIN, for example, a protocol used in treating cancer, has now been introduced to LLM alongside some new AI technology. [12], [13], [14], [15], [16], [17], [18], [19], [20], [21], [22], [23], [24]

In this section, we will look at the historical context and recent advancements of LLMs, as well as an overview of their development and growing importance to SCM. We will explore how these models have transformed traditional supply chain operations and explore the technological underpinnings that enable their sophisticated capabilities. This discussion lays the groundwork for understanding the profound impact that LLMs can have on GCS by guiding both current practices and future innovations.

## 1-1- Historical Background

The development of LLMs in SCM has been marked by significant technological advances and innovative applications. Understanding the historical context and recent advances provides an understanding of how LLMs have transformed GCS and the future potential it holds.

**Early applications of AI in SCM:** The use of artificial intelligence in supply chain management dates back to the late 20th century, when basic AI models were used for inventory management, demand forecasting, and optimization tasks. These early systems relied on rule-based algorithms and statistical methods, which were efficient to some extent, but lacked the adaptability and depth of modern AI technologies.

**Adoption of machine learning** : In the early 2000s, machine learning techniques were integrated into SCM. Machine learning models, capable of learning from data and improving over time, offered more sophisticated and accurate solutions to complex supply chain problems. However, these models required significant amounts of labeled data and were limited in their ability to process unstructured data such as text.

**The emergence of LLMs**: The introduction of transformer models by Vaswani et al. (2017) marked an important milestone in the development of LLMs. The Transformer architecture, with its attention mechanisms, has enabled better management of data and sequential contexts, resulting in significant improvements in natural language understanding and generation. Models such as BERT (Devlin et al., 2018) and GPT-3 (Brown et al., 2020) have demonstrated unprecedented capabilities in language processing and have paved the way for their application in various fields, including SCM. [25], [26], [27]

## 1-2- Technological foundations of large language models

The application of LLMs in SCM is based on a robust technological foundation that allows these models to understand, process, and generate human-like text with high accuracy. Understanding these foundational technologies is critical to understanding how LLMs can be used effectively in SCM to promote efficiency and innovation.

LLMs have revolutionized various fields, including SCM, largely due to their underlying architecture – the transformer – and the powerful attention mechanisms it employs. These innovations have enabled LLMs to efficiently process large amounts of data, making them particularly suitable for complex SCM tasks.

### Transformer Architectures

The Transformer architecture, introduced by Vaswani et al. (2017), marked a significant shift in the way natural language processing models are designed. Unlike previous models that relied heavily on recurring or convolutional layers, the Transformer uses self-attention mechanisms to process data. This architecture is particularly advantageous for processing sequential data, such as text, as it allows for the parallelization of data processing, greatly improving computational efficiency and performance. [25]

**Key components of the Transformer architecture** :

- **Encoder-decoder structure** : The transformer consists of an encoder that processes the input data and a decoder that produces the output. Each encoder and decoder consists of multiple layers of attention mechanisms and neural networks.
- **Self-attention mechanism** : This mechanism allows the model to weigh the meaning of different words in a sentence, regardless of their position. This is essential for understanding context and relationships in language data.
- **Position coding** : Since the transformer does not process data sequentially, position encodings are added to the input embeddings to provide information about the position of each word in the sequence.

### Attention mechanisms

Attention mechanisms are at the heart of the Transformer's architecture. They allow the model to focus on the relevant parts of the input data during processing, which is essential for tasks that require understanding context and relationships over long distances in the data.

**Types of attention mechanisms** :

- **Scaled Dot-Product Attention:** This type of attention calculates the importance of each word in a sequence relative to other words. For this purpose, a weighted sum of the values is calculated, with the weights determined by the query point product and key vectors.
- **Multi-Head Attention** : Instead of using a single attention function, Multi-Head Attention applies multiple attention mechanisms in parallel, each with different linear transformations of the input. This allows the model to capture various aspects of the relationships between words.

## 1-3- Pre-training and fine-tuning techniques

LLMs such as GPT-4 and BERT have revolutionized the field of NLP and are increasingly being applied to SCM. The success of these models depends largely on two critical processes: pre-training and development. These techniques allow models to learn and adapt to specific tasks within the SCM, providing valuable insights and optimizations.

### Pre-workout techniques

**Pre-training for large-scale data** : LLMs are pre-trained with huge datasets containing diverse and comprehensive textual information. The pre-training phase is about unsupervised learning, where the model learns to predict the next word in a sentence, taking into account the context of the previous words. This method, known as masked language modeling for models like BERT or autoregressive modeling for GPT-4, helps the model understand the structure and nuances of human language.

**Transfer Learning**: Pre-trained LLMs take knowledge from large datasets and apply it to specific areas such as SCM. Transfer learning allows these models to generalize their understanding and adapt to different contexts without having to start from scratch for each new task. This makes LLMs particularly effective in addressing GCS-related challenges. [27]

**Domain-specific corpus**: For SCM, pre-training can also include the use of domain-specific corpora including industry reports, supply chain case studies, transaction records, and market analysis. This focused approach helps the model develop a deeper understanding of SCM-specific language and concepts, resulting in more accurate and relevant information. [28]

### Development techniques

**Monitored fine-tuning**: Fine-tuning is the process of training the LLM pre-trained for a specific task with labeled data. In the context of SCM, this could involve refining the model on datasets related to demand forecasting, inventory management, supplier valuation, and logistics optimization. Supervised fine-tuning helps the model learn the specific patterns and requirements of the task, improving performance and accuracy. [26]

**Learning by a few moves and by zero-moves**: Advanced LLMs like GPT-3 and GPT-4 are able to learn with little to no shots, which means they can perform tasks with very few or even no specific examples. This feature is especially useful in SCM, where obtaining annotated data can be rare or expensive. The one-shot learning allows the model to adapt quickly to new tasks and with minimal data input, making it highly flexible and efficient. [27]

**Reinforcement Learning**: Reinforcement learning refines LLMs through a feedback loop in which the model's predictions are continuously refined based on the results of its actions. In SCM, this technique can be applied to optimize decision-making processes such as route planning and inventory management by continuously learning from real-time data and feedback. [29]

## 2- Applications of LLMs in Supply Chain Management

LLMs have had a significant impact on various areas of SCM, providing advanced solutions that improve efficiency, accuracy, and decision-making processes. By leveraging their natural language processing and data analytics capabilities, LLMs can overcome complex challenges and optimize operations in various facets of SCM.

In this section, we will explore the specific applications of LLMs in SCM, focusing on three key areas: demand forecasting and inventory management, supplier relationship management, and logistics and transportation optimization. Each of these areas benefits from the unique strengths of LLMs, allowing organizations to streamline processes, improve collaboration, and effectively mitigate risk.

## 2-1- Demand forecasting and inventory management

Demand forecasting and inventory management are important elements of GCS. Accurate demand forecasting allows businesses to maintain optimal inventory levels and reduce costs associated with overstocks and stock-outs. Integrating LLMs into these processes offers significant improvements in accuracy and efficiency by leveraging advanced data processing capabilities to provide deeper insights and more reliable predictions.

### Demand forecasting

**Increased accuracy through data integration**: LLMs can process and analyze vast data sets, including historical sales data, market trends, economic indicators, and even unstructured data such as social media posts and news articles. This comprehensive analysis allows for more accurate demand forecasts. For example, Li et al. (2023) showed how LLMs can integrate different data sources to improve forecast accuracy and help companies anticipate fluctuations in demand more effectively. [28]

**Real-time demand forecasting**: The ability of LLMs to process data in real-time allows for dynamic demand forecasting. This is especially useful in industries where demand can be highly volatile, such as fashion or electronics. By continually updating forecasts based on the latest data, businesses can respond faster to market changes, optimize inventory, and reduce waste. [30]

**Predictive Analytics and Scenario Planning**: LLMs can be used for predictive analytics, allowing organizations to run different scenarios and evaluate potential outcomes. This capability helps businesses plan for different market conditions and develop risk mitigation strategies. By simulating various demand scenarios, LLMs help supply chain managers make informed decisions about production, procurement, and logistics. [31]

### Inventory

**Optimizing stock levels**: Inventory management is about maintaining the right balance between supply and demand. LLMs can analyze historical inventory data, sales patterns, and external factors to optimize inventory levels. This ensures that businesses have sufficient inventory to meet demand without overstocking that ties up capital and increases inventory costs. [32]

**Automated replenishment**: LLMs can automate the inventory replenishment process by predicting when inventory levels will fall below a certain threshold and generating orders accordingly. This automation reduces the manual effort involved in inventory management and helps maintain continuous supply chain operations. [29]

**Inventory turnover and reduced inventory carrying costs**: By providing accurate demand forecasts and optimizing inventory levels, LLMs help improve inventory turnover rates. A faster turnaround time means products spend less time in the warehouse, reducing storage costs and the risk of obsolescence. This is especially important for perishable goods or products with a short life cycle. [28]

## 2-2- Supplier relationship management

Supplier Relationship Management (SRM) is a critical aspect of SCM that focuses on the interactions between a company and its suppliers. Effective SRM can lead to efficiencies, cost savings, and stronger partnerships. LLMs have become powerful tools for improving SRM by providing advanced analytical capabilities and enabling better communication and decision-making.

### Improve communication

**Automated communication**: LLMs can streamline communication with vendors by automating routine interactions. For example, LLMs can process requests, provide updates on order status, and negotiate terms, freeing up human resources for more strategic tasks. This automation ensures fast and consistent communication, which is essential for maintaining strong relationships with suppliers. [29]

**Natural Language Processing (NLP) Capabilities**: The NLP capabilities of LLMs help analyze communication patterns and sentiment in supplier interactions. By understanding the tone and context of communication, businesses can assess the state of their supplier relationships and proactively resolve potential issues. [32]

### Supplier Performance Analysis

**Data-driven insights** : LLMs can analyze large data sets to evaluate vendor performance. This includes the evaluation of key figures such as delivery times, quality of goods and compliance with contractual conditions. By providing data-driven insights, LLMs help organizations identify which vendors are performing best and which ones may need improvement (Li et al., 2023).

**Predictive Analytics**: LLMs can use historical data to predict future vendor performance. For example, by analyzing past performance trends, LLMs can predict potential delays or quality issues, allowing companies to take preventative steps to mitigate risk. This predictive capability increases the reliability of supply chains. [31]

### Risk Management

**Early warning systems**: LLMs can monitor various data sources, including news, social media, and financial reports, to identify early signs of potential supply chain disruptions. By identifying risks such as financial instability or geopolitical events that may impact suppliers, LLMs provide early warning and allow companies to develop contingency plans. [30]

**Scenario planning**: LLMs facilitate scenario planning by simulating different risk scenarios and their potential impact on the supply chain. This helps businesses prepare for various eventualities and develop strategies to effectively manage supplier risks. [32]

### Improving cooperation with suppliers

**Collaborative platforms**: LLMs can support collaborative platforms that facilitate real-time information sharing between companies and their suppliers. These platforms can include features such as shared dashboards, automated updates, and collaborative scheduling tools. By improving transparency and collaboration, LLMs help build stronger, more resilient relationships with suppliers. [29]

**Contract management**: LLMs can support contract management by analyzing contract terms, monitoring compliance, and suggesting improvements based on historical data. This helps to ensure that all parties abide by the agreed terms and helps negotiate better contracts in the future. [33]

## 2-3- Optimization of logistics and transport

Logistics and transport are important components of SCM, which include the transport and storage of goods from supplier to customer. Efficient logistics and transportation systems reduce costs, improve customer satisfaction, and improve overall supply chain performance. LLMs have the potential to significantly optimize these processes through advanced data analytics, real-time decision-making, and predictive capabilities.

### Route optimization

**Real-time traffic and weather analysis**: LLMs can analyze real-time traffic data and weather forecasts to optimize transportation routes. By processing huge amounts of data from GPS, traffic reports, and weather stations, LLMs can suggest the most efficient routes and avoid delays caused by traffic jams or bad weather conditions. This translates into reduced fuel consumption and faster delivery times. [30]

**Dynamic routing adjustments**: In logistics, conditions can change quickly, requiring dynamic route adjustments. LLMs equipped with real-time data feeds can reroute vehicles on the fly to avoid disruptions such as accidents or road closures. This adaptability increases the reliability and efficiency of the transmission system. [29]

### Fleet Management

**Predictive maintenance**: LLMs can predict the maintenance needs of transportation fleets by analyzing data from vehicle sensors, maintenance logs, and usage patterns. Predictive maintenance helps reduce downtime, extend vehicle life, and reduce maintenance costs by fixing problems before they lead to breakdowns. [28]

**Optimization of fleet utilization**: By analyzing patterns in shipment data and fleet utilization, LLMs can optimize vehicle allocation and scheduling. This ensures efficient fleet utilization, reduces downtime and maximizes return on investment in transportation. [31]

### Optimization of use

**Efficient load planning**: LLMs can optimize load planning by analyzing shipment sizes, weights, and delivery schedules. Efficient load planning ensures full vehicle utilization, reduces the number of trips required and thus reduces transport costs. It also helps to reduce the environmental impact by reducing fuel consumption. [32]

**Multimodal transport planning**: For complex supply chains that involve multiple modes of transportation (e.g., truck, rail, ship), LLMs can plan and optimize the entire transportation process. By integrating data from different modes of transport, LLMs ensure smooth transitions and optimize schedules to reduce delays and costs. [33]

### Optimization of warehouses and distribution centers

**Inventory placement and picking**: LLMs make it possible to optimize the placement of stock within warehouses in order to minimize the distance traveled during picking. By analyzing order patterns and inventory turnover rates, these patterns can suggest optimal storage locations, increasing the efficiency of warehouse operations. [29]

**Automated sorting and shipping**: In distribution centers, LLMs can automate the sorting and routing of packages, ensuring that shipments are sent to the right destinations with minimal effort. This reduces errors, speeds up processing times, and reduces labor costs. [30]

### Supply chain visibility and coordination

**End-to-end visibility**: LLMs provide end-to-end visibility across the entire supply chain by integrating data from multiple sources, including suppliers, carriers, and customers. This visibility allows for better coordination and decision-making, which helps businesses respond quickly to changes in demand or supply. [28]

**Collaborative planning**: LLMs facilitate collaborative planning between different stakeholders in the supply chain. By providing a common platform for data and analytics, LLMs help suppliers, manufacturers, and logistics providers collaborate more effectively and ensure the entire supply chain runs smoothly. [31]

## 3- Improve SCM decision-making with LLMs

The ability to make informed and timely decisions is crucial in the GCS. LLMs have revolutionized decision support systems by providing advanced predictive capabilities and real-time data analytics. These models can process huge amounts of data from multiple sources, generate actionable insights, and help make strategic decisions that improve the efficiency and resilience of supply chains.

In this section, we will explore the application of LLMs in various decision support and prediction scenarios within SCM. We'll look at how LLMs enable real-time data analytics, predictive insights, scenario planning, risk management, and automated decision support systems, and illustrate their transformative impact on supply chain operations.

### 3-1- Real-time data analysis and predictive insights

In SCM's dynamic and complex environment, the ability to analyze data in real-time and generate predictive insights is critical. LLMs have become powerful tools that allow organizations to leverage the vast amounts of data generated in supply chains and provide accurate and timely insights to improve decision-making and operational efficiency.

### Real-time data analysis

**Integration of various data sources**: LLMs can integrate data from a variety of sources, including IoT sensors, ERP systems, CRM platforms, and external data feeds such as market trends and news. This integration provides a complete view of the supply chain and enables more informed decision-making. By processing data in real-time, LLMs help supply chain managers respond quickly to changes and disruptions. [29]

**Improved monitoring and reporting**: Real-time data analysis allows for continuous monitoring of supply chain activity. LLMs can create real-time reports and dashboards that highlight key performance indicators (KPIs) and alert managers to potential issues before they escalate. For example, monitoring inventory levels, production rates, and shipment status can help identify bottlenecks and inefficiencies in a timely manner. [30]

**Anomaly Detection**: LLMs are capable of detecting anomalies in large data sets. By analyzing patterns and trends, these patterns can identify discrepancies that may indicate issues such as equipment failures, supply shortages, or spikes in demand. Early detection of anomalies allows proactive measures to be taken to mitigate risks and keep operations running smoothly. [32]

### Predictive insights

**Demand forecasting**: LLMs improve demand forecasting by analyzing historical sales data, market conditions, and external factors such as economic indicators and seasonal trends. Predictive models can more accurately predict future demand, helping companies optimize inventory levels and production schedules. This reduces the risk of overstocking or stock-outs and allows supply and demand to be effectively adjusted. [28]

**Predictive maintenance**: Predictive maintenance is crucial for production and logistics operations. LLMs analyze machine and vehicle sensor data to predict when maintenance is needed, prevent unexpected failures, and minimize downtime. This predictive capability extends asset life and improves overall operational efficiency. [31]

**Risk Management**: LLMs provide predictive insights for risk management by analyzing data from a variety of sources to identify potential risks. For example, they can predict supply chain disruptions due to geopolitical events, natural disasters, or financial instability of suppliers. This predictive information allows organizations to develop contingency plans and mitigate risks proactively. [33]

**Inventory optimization**: By forecasting demand and analyzing inventory turnover rates, LLMs help optimize inventory levels. Predictive insights ensure that inventory is kept at optimal levels, reducing inventory costs and improving cash flow. This is especially beneficial for industries where perishables or inventory turnover rates are high. [29]

## 3-2- Scenario planning and risk management

Scenario planning and risk management are critical aspects of GCS that allow organizations to anticipate potential disruptions and develop strategies to mitigate them. LLMs have introduced advanced capabilities in this area, leveraging huge amounts of data and sophisticated analytical techniques to predict risk and assist in scenario planning.

### Scenario planning

**Predictive analytics for scenario planning**: LLMs can analyze historical data, market trends, and external factors to predict future scenarios. By simulating different scenarios, businesses can prepare for various potential outcomes and develop solid strategies. For example, LLMs can predict how changes in consumer behavior, geopolitical events, or natural disasters may affect the supply chain. [28]

**What-if analysis**: LLMs facilitate what-if analysis by assessing the impact of various variables on the supply chain. This allows companies to understand the consequences of various environmental decisions and changes. For example, companies can assess the impact of rising tariffs on raw materials or the impact of a supplier's departure. [30]

**Optimization of emergency plans**: Based on scenario analysis, LLMs help companies develop and optimize contingency plans. By identifying the most effective responses to potential disruptions, LLMs enable organizations to minimize downtime and maintain continuity. These include alternative sourcing strategies, inventory reserves, and flexible logistics arrangements. [31]

### Risk Management

**Risk identification and assessment**: LLMs can scan and analyze data from a variety of sources, including news articles, social media, financial reports, and supplier performance records, to identify potential risks. This comprehensive risk identification process allows organizations to be aware of new threats and assess their potential impact on the supply chain. [32]

**Real-time risk monitoring**: By providing real-time monitoring capabilities, LLMs allow supply chain managers to keep tabs on risks as they evolve. This includes monitoring geopolitical events, natural disasters, market fluctuations, and other factors that could disrupt the supply chain. Real-time alerts allow for rapid response to mitigate these risks. [29]

**Predictive Risk Modeling**: LLMs use predictive risk modeling to predict the likelihood and impact of various risks. This includes analyzing trends in historical data and identifying indicators that precede disruptions. For example, LLMs can predict vendor financial instability by analyzing payment trends and market conditions, allowing businesses to take proactive action. [33]

**Supply Chain Resilience**: LLMs help build supply chain resilience by highlighting vulnerabilities and suggesting improvements. These include recommending supplier diversification, increasing inventory reserves, and improving logistics capabilities. By strengthening the supply chain, businesses can better withstand and recover from disruptions. [31]

## 3-3- Automated decision support systems

SCM's automated decision support systems (ADS) use state-of-the-art technologies to help managers make informed decisions. LLMs are now an integral part of these systems, offering sophisticated data analytics, predictive modeling, and real-time decision-making capabilities. This section explores how LLMs enhance automated SSD in SCM, with a focus on recent advances and practical applications.

### Improving Decision-Making with LLMs

**Data Integration and Analysis**: LLMs can process and integrate data from a variety of sources, including internal databases, market reports, social media, and IoT sensors. This capability enables comprehensive analysis and provides a holistic view of the supply chain. By analyzing this data, LLMs can identify patterns and trends that influence strategic decisions. [28]

**Predictive Modeling**: LLMs excel at predictive modeling, which is essential for anticipating future events and trends. These models can forecast demand, predict potential disruptions, and estimate lead times with high accuracy. Predictive insights enable managers to make proactive decisions and reduce the risk of stock-outs and overstocks. [29]

**Real-time decision support**: The ability of LLMs to analyze data in real-time ensures that decision-makers have access to the most up-to-date information. Real-time analytics allow for immediate response to changing conditions, such as changes in demand or supply chain disruptions. This responsiveness is essential to maintain efficiency and minimize delays. [30]

### Supply Chain Management Applications

**Inventory**: Automated DSS based on LLMs can optimize inventory levels by predicting future demand and suggesting reorder points. This helps to maintain optimal stock levels, reduce inventory costs, and prevent stock-outs. For example, a retail company can use LLMs to analyze sales data and predict seasonal

fluctuations in demand to ensure that inventory is adjusted accordingly. [28]

**Supplier selection and management**: LLMs can evaluate supplier performance based on a variety of criteria, such as delivery times, quality of goods, and reliability. The automated DSS can rank vendors and suggest the best options based on this analysis. This improves supplier selection processes and supplier relationship management. [31]

**Logistics and transport planning**: In logistics, LLMs help with route optimization, freight planning, and fleet management. Automated DSS can analyze traffic patterns, weather conditions, and delivery schedules to suggest the most efficient routes. This reduces transport costs and delivery times. [32]

**Risk Management**: LLMs enable the automated DSS to assess and mitigate risk by analyzing data from multiple sources, including geopolitical events, economic indicators, and supplier performance. Predictive analytics can predict potential risks and allow businesses to develop contingency plans and maintain supply chain resilience. [33]

**Production planning**: In the manufacturing sector, LLMs support production planning by predicting the demand for finished goods and raw materials. Automated DSS can suggest production schedules that align with demand forecasts, ensuring that resources are used efficiently and that production meets market demands. [29]

# 4- Customization and customization in SCM

In the ever-changing SCM landscape, the ability to tailor solutions to specific industry needs and adapt to dynamic environments is essential. LLMs have become powerful tools that can be adapted to meet the unique challenges of different industries. These models provide adaptive learning capabilities that allow supply chains to remain resilient and adapt to changing conditions.

This section examines two critical aspects of using LLMs in SCM: tailoring supply chain solutions to specific industries and implementing adaptive learning for dynamic supply chain environments. By exploring these areas, we aim to show how LLMs can increase the efficiency and effectiveness of supply chains in different contexts.

## 4-1- Tailored supply chain solutions for specific industries

In SCM's diverse landscape, one-size-fits-all solutions are often not enough to meet the unique challenges and requirements of different industries. LLMs offer a transformative approach by providing customized supply chain solutions tailored to the specific needs of different industries. These models leverage huge amounts of industry-specific data, advanced analytics, and real-time processing to provide tailored strategies that increase efficiency, reduce costs, and improve overall performance.

### Applications specific to the LLM industry

**Automobile industry**: The automotive industry is characterized by complex global supply chains that require precise coordination and on-time delivery of parts and materials. LLMs help optimize production schedules, manage supplier relationships, and predict fluctuations in demand. By analyzing data from a variety of sources, including manufacturing processes, market trends, and supplier performance, LLMs provide actionable insights that streamline operations and improve supply chain resilience. [30]

**Health & Pharmacy**: In the healthcare and pharmaceutical sectors, supply chain reliability is critical due to the life-saving nature of the products involved. LLMs help accurately forecast demand for medical supplies, streamline inventory management, and ensure timely delivery of medicines and vaccines. These models can also track regulatory compliance and manage risks related to supply chain disruptions to ensure that critical products are always available when needed. [28]

**Retail & E-commerce**: The retail and e-commerce industries face unique challenges, such as managing large volumes of SKUs, seasonal fluctuations in demand, and last-mile delivery logistics. LLMs improve these operations by providing real-time inventory management, personalized marketing strategies, and streamlined logistics planning. By integrating data from customer behavior, sales trends, and logistics constraints, LLMs help retailers maintain optimal inventory levels, shorten delivery times, and improve customer satisfaction. [29]

**Food and Beverage**: The food and beverage industry requires efficient management of perishable goods, strict adherence to safety standards and on-time distribution. LLMs can predict demand based on factors such as weather, seasonal trends, and consumer preferences. These models also ensure compliance with food safety regulations and optimize the supply chain to minimize waste and improve freshness, improving profitability and customer confidence. [31]

**High-tech electronics and manufacturing**: The high-tech electronics and manufacturing industry faces rapid product lifecycles, complex assembly processes, and significant supply chain risks. LLMs help predict demand for new products, manage multi-tiered supplier networks, and optimize production schedules. By analyzing technology trends, market demand, and supply chain vulnerabilities, LLMs help these industries stay ahead of the competition and respond quickly to market changes. [32]

## Adapting LLMs to the needs of the industry

**Domain-specific training data**: To effectively adapt LLMs to different industries, it is imperative to use domain-specific training data. This data includes industry reports, technical documentation, regulatory guidance, and historical performance data. Aligning LLMs with this data allows them to understand the specific nuances and challenges of each industry, resulting in more accurate and relevant insights. [33]

**Fine-tuning and customization**: Adapting LLMs to industry-specific tasks improves their performance. In the pharmaceutical industry, for example, models can be refined to predict drug shortages or optimize clinical trial logistics. This process involves adjusting the model parameters to focus on the most critical aspects of the industry and ensuring that the results are both accurate and actionable. [29]

**Integration with industry-specific systems**: LLMs can be integrated with existing industry systems such as enterprise resource planning (ERP), SCM software, and customer relationship management (CRM) platforms. This integration enables a continuous flow of data and real-time updates, ensuring that decision-makers have access to the most up-to-date and relevant information. By working with these systems, LLMs improve their capabilities and provide more comprehensive solutions. [28]

## 4-2- Adaptive Learning for Dynamic Supply Chain Environments

The dynamic nature of supply chain environments requires systems that can learn and adapt quickly to changing conditions. LLMs equipped with adaptive learning capabilities offer significant advantages in managing these complexities. Adaptive learning enables LLMs to continuously improve their performance by incorporating new data and feedback to ensure efficient and resilient supply chain operations.

### Key Components of Adaptive Learning in LLMs

**Continuous Data Integration**: Adaptive learning systems rely on the continuous integration of data from various sources such as market trends, consumer behavior, supplier performance, and environmental conditions. By continuously ingesting and processing new data, LLMs can update their models in real-time, providing the most up-to-date insights and recommendations. [28]

**Real-time feedback loops**: LLMs with adaptive learning skills use real-time feedback loops to refine their predictions and decision-making processes. These feedback loops consist of constantly comparing the predicted results with the actual results, allowing the model to adjust its parameters and improve its accuracy over time. [30]

**Self-learning algorithms**: Self-learning algorithms allow LLMs to autonomously improve their understanding of supply chain dynamics. These algorithms allow models to detect patterns, detect anomalies, and predict future events without the need for explicit programming for each scenario. This capability is critical to adapting to unforeseen changes and maintaining operational efficiency. [33]

## Applications in dynamic supply chain environments

**Demand forecasting**: In dynamic supply chains, accurate demand forecasting is essential. Adaptive learning allows LLMs to continuously refine their demand forecasts by integrating real-time sales data, market trends, and external factors such as economic conditions and consumer preferences. This ensures that stock levels are optimized and that stock-outs and overstocks are reduced. [31]

**Supplier management**: Adaptive learning improves supplier management by providing real-time insights into supplier performance and potential risks. LLMs can analyze data from supplier interactions, delivery records, and market conditions to predict supplier reliability and recommend adjustments to supplier strategies. This helps to maintain a robust and responsive supply chain. [28]

**Optimization of logistics**: The dynamics of logistics operations, including route planning and fleet management, benefit significantly from adaptive learning. LLMs can optimize logistics by analyzing real-time traffic data, weather conditions, and delivery schedules. This allows for dynamic route adjustments and efficient use of resources, reducing delivery times and costs. [29]

**Risk Management**: Adaptive learning in LLMs improves risk management by continuously monitoring potential disruptions. By integrating data from geopolitical events, natural disasters, and financial markets, LLMs can predict risks and propose proactive measures to mitigate their impacts. This helps ensure that supply chains remain resilient and can recover quickly from disruptions. [28]

**Inventory**: In dynamic environments, maintaining optimal inventory levels is a challenge. Adaptive learning allows LLMs to accurately predict inventory demand by analyzing sales patterns, supplier delivery times, and market conditions. This helps reduce inventory costs and improve service levels by ensuring the right products are available at the right time. [30]

# 5- Challenges and ethical considerations

Implementing LLM in SCM offers transformative potential, increases efficiency, and streamlines operations. However, the ethical considerations associated with the use of these cutting-edge technologies are paramount. To ensure that LLMs are used responsibly, important concerns such as bias, fairness, privacy, security, intelligibility, and transparency must be addressed.

In this section, we will explore these critical ethical issues in detail. We will discuss how to address bias and ensure fairness, ensure data privacy and security, and maintain intelligibility and transparency in decision-making processes. Each of these topics is critical to building trust and accountability when using LLMs in SCM.

## 5-1- Addressing bias and ensuring fairness

In the SCM, the use of LLMs has introduced new efficiencies and capabilities. However, the issue of bias and fairness in these models is a major issue. Ensuring that LLMs operate without inherent bias and that their outcomes are fair and equitable is critical to maintaining trust and integrity in the supply chain.

### Understanding bias in LLMs

**Causes of bias**: Bias in LLMs can come from a variety of sources, including the data used for training, the algorithms used, and the implementation processes. Training data may contain historical biases or reflect societal inequities that may be inadvertently learned by models. In addition, algorithmic biases can result from model design and optimization criteria. [34]

**Types of distortions:**

- **Data bias** : Occurs when training data is not representative of the actual diversity modeled. This can lead to biased predictions and decisions that favor certain groups over others.

- Algorithmic bias : Stems from the architecture and the learning process of the model, which can prioritize certain models or features, resulting in biased results.
- Human bias : Introduced during the development, deployment, and evaluation phases by the people involved in these processes.

### Ensuring fairness

**Diverse and representative training data**: To avoid data bias, it is important to use diverse and representative datasets for LLM training. The goal is to collect data from different sources and ensure that different demographics and scenarios are adequately represented. Efforts should be made to identify and correct underrepresented areas in the data. [33]

**Bias detection and mitigation techniques** : The implementation of bias detection and mitigation techniques is essential to the development of equitable LLMs. These techniques include:
- **Preprocessing** : Modify the training data to reduce distortion before it is introduced into the model. This could be balancing the dataset or removing distorted entries.
- **Processing** : Incorporate fairness constraints and regularization techniques during the model training process to ensure fair treatment of all data points.
- **Post-processing** : Adjust the model outputs to correct detected distortions. This may involve recalibrating predictions or applying fairness measures to evaluate and adjust the results.

**Fairness measures**: The development and application of equity measures is essential to assess the performance of LLMs in GCS. Indicators such as demographic parity, equal opportunity, and different impact can help assess whether the model's decisions are fair to different groups. Regular audits and evaluations against these metrics are necessary to maintain fairness. [35]

**Transparent and explainable AI**: Transparency and explainability are crucial to address bias and ensure fairness in the LLM. Models should be designed to provide information about their decision-making processes. Explainable AI techniques such as SHAP (SHAP values) or LIME (Local Interpretable Model-agnostic Explanations) can help stakeholders understand how the model arrived at certain decisions and identify potential biases. [36]

**Ethical AI Practices**: Adopting ethical AI practices involves creating guidelines and standards for the development and deployment of LLMs in SCM. This includes adhering to principles such as accountability, transparency and fairness. Companies should set up ethics review committees to monitor the use of AI technologies and ensure compliance with these standards. [37]

## 5-2- Privacy and security concerns

In the age of digital transformation, SCM is increasingly relying on cutting-edge technologies such as LLMs to streamline operations and improve decision-making. However, the integration of these technologies raises significant privacy and security concerns. Ensuring the protection of sensitive information while leveraging the capabilities of LLMs is critical to maintaining trust and compliance in SCM.

### Understanding Privacy and Security in LLMs

**Type of data in SCM** Supply chains generate large amounts of data, including sensitive information such as supplier contracts, customer data, and financial transactions. This data is critical to efficient operations, but it also poses significant risks to data privacy if not properly managed. [38]

**Data processing and storage**: LLMs require extensive data for training and debugging, which often requires the processing and storage of large data sets. It is essential to ensure that this data is handled securely and in compliance with data protection regulations to prevent unauthorized access and breaches. [33]

### Significant privacy and security concerns

**Data Breaches**: Data breaches can occur when unauthorized individuals gain access to confidential information. This can lead to significant financial losses, reputational damage, and legal ramifications. Protecting supply chain data from security breaches is a major concern when deploying LLMs. [31]

**Misuse of data**: LLMs can inadvertently learn and propagate biases in training data, resulting in misuse of sensitive information. Ensuring that data is used ethically and responsibly is critical to maintaining the integrity of supply chain operations. [28]

**Compliance**: Various regulations, such as the General Data Protection Regulation (GDPR) and the California Consumer Privacy Act (CCPA), have strict data privacy and security requirements. Compliance with these regulations is critical for global businesses, and LLMs must be designed to meet these regulatory standards. [32]

**Access Control**: Effective access control mechanisms are needed to ensure that only authorized personnel can access sensitive data. Implementing robust authentication and authorization processes helps ensure that unauthorized individuals can access or modify data. [30]

### Strategies to improve data privacy and security

**Data encryption**: Encrypting data at rest and in transit is a basic security measure that protects sensitive information from unauthorized access. Advanced encryption techniques ensure that data is safe, even if it is intercepted or accessed by malicious actors. [29]

**Secure data storage**: Storing data in secure environments, such as cloud services with strong security protocols, helps protect against security breaches. Regular security reviews and updates of storage systems ensure continuous protection against new threats. [33]

**Anonymization and depersonalization**: Anonymizing and anonymizing data before using it to train LLMs reduces the risk of exposing sensitive information. This process involves removing or masking personal identifiers to ensure that individuals cannot be easily identified from the data. [28]

**Compliance Frameworks**: Developing and adhering to comprehensive compliance frameworks ensures that privacy and security measures meet regulatory standards. Regular audits and assessments help maintain compliance and identify areas for improvement. [31]

**Robust access controls**: Implementing multi-factor authentication (MFA) and role-based access control (RBAC) enhances security by ensuring that only authorized users can access sensitive data. These measures help prevent unauthorized access and ensure that users have the appropriate level of access based on their role. [30]

## 5-3- Intelligibility and transparency in decision-making

In SCM, the introduction of LLMs introduced advanced capabilities for data analysis, forecasting, and decision support. However, the complexity of these models often makes their decision-making processes opaque. Ensuring the intelligibility and transparency of these decisions is critical to building trust, enabling compliance, and improving the efficiency of supply chain operations.

### Importance of intelligibility and transparency

**Trust and responsibility**: Intelligibility and transparency are key to building trust among stakeholders. When decision-making processes are clear and understandable, stakeholders are more likely to trust and rely on the information gained from LLMs. This trust is crucial for the successful implementation and introduction of AI-based solutions in SCM. [36]

**Compliance**: Many industries are subject to regulations that require transparency in automated decision-making. Ensuring that LLMs comply with these regulations helps companies avoid legal risks and penalties. Transparency of AI systems is also necessary to meet ethical standards and uphold corporate social responsibility. [37]

**Improved quality of decisions**: Understanding how LLMs arrive at their decisions allows supply chain managers to assess the quality and appropriateness of those decisions. It helps identify potential biases or errors and refine models for better accuracy and efficiency. Transparent models enable the continuous improvement and optimization of supply chain processes. [33]

### Techniques to improve intelligibility and transparency

**Explainable AI (XAI) methods:** Explainable AI (XAI) techniques aim to make the decision-making processes of LLMs more understandable. Methods such as SHapley Additive exPlanations (SHAP) and Local Interpretable Model-agnostic Explanations (LIME) provide insight into the features that most influence model predictions. These techniques help stakeholders understand the factors that influence model decisions. [39]

**Model documentation and reporting**: Creating comprehensive documentation and reporting mechanisms for LLM is crucial for transparency. This includes detailed description of data sources, preprocessing steps, model architectures, and decision criteria. Model maps and dataset datasheets are tools that can be used to effectively document and communicate these aspects. [35]

**Interactive visualization tools**: Interactive visualization tools allow users to explore the data and decision-making process of LLMs. These tools can show how different inputs affect the model's outputs, allowing users to understand the relationships between variables and the resulting decisions. Visualization makes complex models more accessible and interpretable. [40]

**Testing and validation**: Regular audits and validations of LLMs ensure that the models are working as intended and comply with transparency standards. This includes checking model performance, checking for bias, and validating the accuracy of predictions. Independent audits provide an additional level of accountability and trust. [41]

## 6- Case Studies and Real-World Implementations

LLMs have profoundly influenced SCM by improving data processing, predictive analytics, and decision-making skills. Their integration into SCM has led to significant improvements in various industries, from automotive to healthcare, streamlining operations, reducing costs, and increasing efficiency.

In this section, we will look at the real-world impact of LLMs on GCS through success stories and case studies. These examples will illustrate how different industries have used the LLM to achieve significant benefits. In addition, we will discuss lessons learned and best practices that have emerged from these implementations, providing valuable insights for companies looking to adopt LLMs in their supply chain processes.

### 6-1- LLM Success Stories in GCS

The integration of LLMs into SCM has led to many success stories that demonstrate the transformative potential of these technologies. These models have enabled companies to increase efficiency, reduce costs, and improve decision-making in various facets of the supply chain. Below are some notable examples from various industries that illustrate how LLMs have been successfully applied to SCM.

**Automotive: Supplier Network Optimization**

In the automotive industry, it is essential to manage a large network of suppliers and ensure on-time delivery of components. An automaker implemented an LLM-based system to analyze supplier performance data, predict potential delays, and optimize procurement strategies. By integrating data from supplier audits, shipping records, and market conditions, the model provided actionable insights that helped improve supplier selection and increase supply chain resilience. As a result, the company saw a 15% reduction in lead times and a 20% reduction in procurement costs. [30]

**Health: Improving pharmaceutical supply chains**

The healthcare sector has faced significant challenges in managing the supply chain of medical supplies and vaccines, especially during the COVID-19 pandemic. A pharmaceutical company used LLMs to forecast demand for essential medicines and optimize inventory levels. The model analyzed data from epidemiological reports, hospital inventories, and production schedules to ensure that supplies were distributed efficiently and shortages were minimized. This approach has not only improved the availability of essential medicines, but has also reduced waste by 10% and shortened delivery times by 25%. [28]

**Retail: Personalizing the customer experience**

In retail, customer satisfaction is intimately linked to the efficiency of the supply chain. A leading e-commerce platform used LLMs to improve its inventory management and personalize the customer experience. By analyzing customer behavior, sales trends, and market data, the model provided real-time recommendations for inventory replenishment and targeted marketing campaigns. This has led to a significant reduction in stock-outs and increased sales during peak periods. The company reported a 30% increase in customer satisfaction and a 25% increase in revenue. [29]

**Agri-food: Reducing food waste**

The food and beverage industry is particularly susceptible to supply chain inefficiencies due to the perishable nature of its products. A major food retailer implemented an LLM-based system to optimize inventory management and reduce food waste. The model analyzed sales data, weather forecasts, and promotional activities to accurately predict demand and adjust inventory levels accordingly. This has resulted in a 15% reduction in food waste and a 12% increase in overall profitability. [31]

**Electronics: Streamlining production and logistics**

In the electronics manufacturing industry, maintaining an efficient supply chain is crucial to meet market demand and control production costs. An electronics manufacturer used LLMs to streamline its production planning and logistics operations. By integrating data from production lines, supplier networks, and market trends, the model optimized production plans and logistics routes. This has resulted in a 20% improvement in production efficiency and a 15% reduction in logistics costs. [32]

## 6-2- Lessons learned and best practices

The integration of LLMs into SCM has resulted in valuable insights and significant improvements in various areas. However, the road to successful implementation has been fraught with pitfalls and learning opportunities. This section describes the key takeaways and best practices that arise from applying LLMs in SCM.

**Teach**

**Importance of data quality and integration**: One of the most important lessons is the importance of high-quality, integrated data. LLMs rely heavily on huge amounts of data to generate accurate predictions and insights. Inconsistent or incomplete data can lead to suboptimal results. Organizations have learned to invest in robust data management systems that ensure data is collected, cleaned, and integrated from multiple sources. [28]

**Continuous training and updating of models**: LLMs must be continuously trained and updated with new data to maintain their accuracy and relevance. The dynamic nature of supply chains requires models that adapt to changing conditions such as market trends, consumer behavior, and geopolitical events. Regular updates and renewed training ensure that LLMs remain effective and provide up-to-date information. [29]

**Balance between complexity and interpretability**: Although LLMs can handle complex tasks, their complexity can sometimes hinder interpretability. Companies have found that it is crucial to find a balance between the sophistication of the models and their interpretability. The use of explainable AI techniques helps stakeholders understand decision-making processes, build trust, and ensure that models are used effectively. [33]

**Ethical Considerations and Avoidance of Bias**: Incorporating ethical considerations and mitigating bias

in the LLM is essential to fair and equitable decision-making. Organizations have learned the importance of implementing bias detection and mitigation strategies to ensure that their models do not perpetuate existing inequities. Ethical AI practices are now a fundamental part of the use of LLMs in SCM. [31]

**Interdepartmental collaboration**: Successful implementation of LLMs often requires collaboration across different departments, including IT, data science, operations, and compliance. Cross-functional teams ensure that all aspects of the supply chain are considered and that models are aligned with the overall business strategy. This collaborative approach leads to more comprehensive and effective solutions. [32]

### Best practices

**Robust data governance framework**: Implementing a robust data governance framework is essential to managing the data lifecycle. This includes establishing clear policies for the collection, storage, access, and use of data. Data governance ensures data integrity and security, providing a solid foundation for the effective operation of LLMs. [30]

**Regular audits and performance monitoring**: Regular audits and performance monitoring help LLMs perform as intended and create value. This includes evaluating the accuracy, efficiency, and fairness of the models. Continuous monitoring allows for timely adjustments and improvements, so models remain aligned with business goals. [28]

**Transparent and explainable templates**: The development of transparent and explainable models is crucial to gaining the trust of stakeholders. With explainable AI tools and techniques, companies can make LLM decision-making processes more understandable. Transparency ensures that users can see how conclusions are drawn and trust the models' recommendations. [29]

**Ethical implementation of AI**: Adopting ethical principles for AI involves implementing practices that promote fairness, accountability, and transparency. This includes conducting bias audits, ensuring diverse training data, and complying with ethical guidelines and regulations. Implementing ethical AI is essential to maintaining the integrity and societal acceptance of LLMs in SCM. [33]

**Scalability and flexibility**: Developing LLMs with scalability and flexibility in mind allows them to adapt to changing supply chain needs. Scalable models can handle increasing volumes of data and more complex tasks, while flexible models can adapt to new data types and changing market conditions. This adaptability is the key to long-term success. [31]

**User training and retention**: For a successful implementation, it is essential to ensure that users are properly trained and familiar with the technology. Training programs should focus on improving users' understanding of the capabilities and limitations of the LLM, as well as how to interpret and respond to lessons learned. Engaged users are more likely to realize the full potential of LLMs. [32]

## 7- Future SCM Trends and Innovations with LLM

The integration of LLMs into SCM has paved the way for significant advancements in efficiency, decision-making, and overall operational performance. As technology advances, emerging trends and innovations promise to continue to transform SCM. Understanding these future trends is crucial for businesses that want to stay competitive and respond to market dynamics.

In this section, we explore the impact of new technologies and the long-term implications of integrating LLMs into SCM. These insights help companies anticipate change, prepare for future challenges, and seize new opportunities for growth and innovation.

### 7-1- New technologies and their impact on GCS

The SCM landscape is constantly evolving, driven by rapid technological advancements. New technologies are changing the way supply chains work, providing new opportunities for efficiency, resilience and innovation. This section explores some of the key emerging technologies and their potential impact on GCS.

### Internet of Things (IoT)

**Impact on SCM:** IoT involves connecting physical devices over the internet, allowing them to collect and share data. In SCM, IoT devices can track inventory levels, monitor the status of goods in transit, and provide real-time supply chain insights.

- **Real-time tracking: IoT sensors** on containers and shipping vehicles enable real-time tracking of goods, allowing for better visibility and inventory management.
- **Condition monitoring:** Sensors can monitor the temperature, humidity, and other conditions of perishables, ensuring quality is maintained throughout the supply chain.
- **Predictive maintenance:** IoT devices can predict when machines and vehicles need maintenance, reducing downtime and improving efficiency. [28]

### Blockchain technology

**Impact on SCM:** Blockchain technology offers a decentralized and secure way to record transactions and track assets. It increases the transparency, traceability and security of SCM.

- **Improved traceability:** Blockchain provides tamper-proof record of the entire supply chain, from raw materials to finished goods, improving traceability and accountability.
- **Fraud prevention:** The immutability of blockchain records helps prevent fraud and counterfeiting in the supply chain.
- **Smart Contracts:** Automated contracts on the blockchain can trigger actions when certain conditions are met, simplifying processes and reducing the need for intermediaries. [29]

### Artificial Intelligence (AI) and Machine Learning (ML)

**Impact on SCM:** AI and ML technologies are transforming SCM by automating complex processes, improving decision-making, and providing predictive insights.

- **Demand forecasting:** AI algorithms can analyze historical sales data and market trends to more accurately predict future demand and help businesses optimize inventory levels.
- **Route optimization:** ML models can optimize delivery routes based on real-time traffic data, weather conditions, and delivery schedules, improving logistics efficiency.
- **Risk management for suppliers:** AI can assess and predict supplier risks by analyzing financial reports, market data, and historical performance, enabling proactive risk management. [31]

### Robotics and Automation

**Impact on SCM:** Robotics and automation are revolutionizing warehousing, production, and distribution processes by increasing efficiency and reducing human error.

- **Automated warehousing:** Robots can take over repetitive tasks such as picking, packing, and sorting, speeding up warehouse operations and reducing labor costs.
- **Production line automation:** Automated production lines improve manufacturing efficiency and consistency and allow for faster responses to changes in demand.
- **Autonomous vehicles:** Autonomous trucks and drones can transport goods with minimal human intervention, reducing delivery times and costs. [32]

### Augmented reality (AR) and virtual reality (VR)

**Impact on SCM:** AR and VR technologies provide immersive and interactive experiences, improving SCM training, planning, and operational efficiency.

- **Advanced training:** AR and VR can simulate real-world scenarios for training purposes and help employees develop their skills in a controlled environment.
- **Warehouse management:** AR can help employees locate and pick items more efficiently by overlaying digital information in the physical world.
- **Virtual prototyping:** Virtual reality allows companies to create virtual prototypes of products and packaging, reducing the time and cost of physical prototyping. [30]

### 5G connectivity

**Impact on SCM:** The advent of 5G technology provides faster and more reliable wireless communication that supports real-time data exchange and advanced IoT applications in SCM.

- **Real-time data transmission:** 5G enables instantaneous data transmission and supports real-time monitoring and decision-making throughout the supply chain.
- **Improved IoT applications:** Increased bandwidth and low latency in 5G networks improve the performance of IoT devices and increase their impact on SCM.
- **Remote Control:** 5G supports remote monitoring and control of supply chain operations, enabling more flexible and resilient supply chain management. [33]

### Edge Computing

**Impact on SCM:** Edge computing processes data closer to the source instead of relying on centralized cloud servers, reducing latency and bandwidth usage.
- **Faster processing:** Edge computing allows data to be processed in real-time at the source, improving response times and decision-making.
- **Reduced bandwidth:** By processing data locally, edge computing reduces the amount of data transmitted to central servers, reducing bandwidth costs.
- **Improved Security:** Keeping data closer to the source can improve security by reducing the risk of interception in transit. [28]

## 7-2- Long-term effects of LLM integration

Integrating LLMs into SCM has already shown significant short-term benefits, including better decision-making, increased efficiency, and greater responsiveness. Moving forward, the long-term effects of the integration of LLM promise to fundamentally reshape the SCM landscape. This section examines these long-term impacts, with a focus on sustainability, innovation, workforce transformation, and competitive advantages.

### Sustainability and environmental impact

**Optimized use of resources** : LLMs allow for more accurate demand forecasting and inventory management, reducing waste and ensuring optimal use of resources. By aligning production with actual demand, companies can minimize overproduction and overstocking, resulting in a lower carbon footprint and more sustainable operations (Li et al., 2023).

**Efficient logistics and transport**: Advanced LLMs can optimize logistics and transportation routes, reducing fuel consumption and emissions. Real-time data analysis enables dynamic adjustments to delivery schedules, minimizes empty runs, and increases overall transportation efficiency. [30]

**Supporting the circular economy**: LLMs can facilitate the transition to a circular economy by improving the tracking and management of reusable and recyclable materials. Improved visibility and predictive analytics help companies effectively manage reverse logistics and promote sustainability practices. [33]

### Innovation and technological progress

**Continuous improvement and adaptation**: The adaptive learning capabilities of LLMs ensure that supply chain processes continuously improve over time. As these models are exposed to new data and scenarios, they refine their algorithms, leading to incremental performance and innovation in SCM practices. [29]

**Integration with new technologies**: LLMs are increasingly integrated with other emerging technologies such as IoT, blockchain, and robotics. This convergence will create smarter, more autonomous supply chains that can make real-time decisions and optimize themselves. [31]

**Advanced Predictive Analytics**: The predictive power of LLMs will continue to evolve, providing deeper insights into market trends, consumer behavior, and potential disruptions. These enhanced predictive analytics will help organizations anticipate and proactively respond to change to maintain a competitive advantage. [28]

### Workforce Transformation

**New Skills Required**: The integration of the LLM into SCM will create a demand for new qualifications and skills. Employees must be proficient in data analytics, AI technologies, and digital tools to realize the full potential of LLMs. Continuous learning and education will become essential. [32]

**Advanced decision-making**: LLMs will improve human decision-making by providing data-driven insights and recommendations. This collaboration between humans and AI will enable more informed and strategic decisions and improve overall supply chain performance. [30]

**Development and job creation**: While some traditional supply chain roles can be automated, new jobs will emerge that will focus on AI-driven process management and optimization. Changing job roles will require a shift in training and workforce development strategies to meet changing demands. [33]

### Competitive Advantage

**Improved agility and resilience**: Companies that effectively integrate LLMs into their supply chain operations will gain significant agility and resilience. The ability to respond quickly to market changes, disruptions, and opportunities provides a competitive advantage in a dynamic business environment. [31]

**Strategic differentiation**: LLMs allow companies to differentiate themselves through better customer service, efficient operations, and innovative solutions. Companies that harness the power of LLMs to optimize their supply chains will be characterized by reliability and efficiency. [29]

**Long-term savings**: The efficiencies achieved through the integration of LLM translate into long-term cost savings. Waste reduction, inventory optimization, and efficient logistics help reduce operating costs and allow companies to reinvest savings in strategic and growth initiatives. [28]

## 8- Discussion

This article explores the integration of LLMs into GCS and highlights its transformative effect on different facets of GCS. Table 1 Overview of the main themes and applications that will be discussed:

*Table 1: The table of keywords of the paper sections*

| Section | Subsection | Key Points |
|---|---|---|
| Introduction to LLMs in SCM (1-) | Historical Background (1-1-) | - The beginnings of AI in SCM: rule-based systems, machine learning. <br> - Evolution with transformer models. <br> - Modern applications and improvements. |
| | Technological Foundations of Large Language Models (1-2-) | - Transformer architecture: self-respect, parallel processing, encoder-decoder structure. <br> - Attention mechanisms: self-attention, multi-headed attention, sparse attention. |
| | Pre-training and fine-tuning techniques (1-3-) | - Large-scale data pre-training, transfer learning. <br> - Domain-specific corpus. <br> - Supervised fine-tuning, learning by a few strokes/zero-strokes, learning by reinforcement. |
| Applications of LLMs in SCM (2-) | Demand forecasting and inventory management (2-1-) | - Improved forecast accuracy, inventory optimization. <br> - Analysis of historical data and market trends. |
| | Supplier Relationship Management (2-2-) | - Improved communication and negotiation. <br> - Insights into supplier performance and risk prediction. |
| | Optimization of logistics and transport (2-3-) | - Route optimization, delay prediction, cost reduction. <br> - Real-time data processing for efficient transport management. |
| Improving SCM decision-making with LLMs (3) | Real-time data analytics and predictive insights (3-1-) | - Integration of various data sources. <br> - Improved monitoring and reporting. <br> - Anomaly detection. |

|  | Scenario Planning and Risk Management (3-2-) | - Predictive analytics for scenario planning.<br>- Simulation analysis.<br>- Optimization of emergency plans. |
|---|---|---|
|  | Automated decision support systems (3-3-) | - Data integration and analysis.<br>- Predictive modeling.<br>- Real-time decision support. |
| Customization and customization in SCM (4-) | Tailor-made supply chain solutions for specific industries (4-1-) | - Applications specific to the LLM industry.<br>- Adaptation of LLMs to industry requirements. |
|  | Adaptive learning for dynamic supply chain environments (4-2-) | - Continuous data integration.<br>- Real-time feedback loops.<br>- Self-learning algorithms. |

### Historical Background and Technological Foundations

Historically, SCM has relied on foundational AI models and machine learning techniques for tasks such as inventory management and demand forecasting. The advent of transformer models, introduced in particular by Vaswani et al. (2017), marked a crucial advance in AI capabilities. These models, with their encoder-decoder structures and attention mechanisms, set the stage for the demanding applications we see today.

### Pre-training and fine-tuning techniques

LLMs benefit from pre-training techniques and large-scale data development. Pre-training with rich datasets allows these models to understand linguistic subtleties, while domain-specific fine-tuning tailors them to specific SCM tasks. Techniques such as supervised fine-tuning, one-shot learning, and reinforcement learning further enhance their adaptability and efficiency.

### Applications in SCM

**Demand forecasting and inventory management** : LLMs improve the accuracy of demand forecasts and optimize inventory levels by integrating various data sources, including historical sales data and market trends. This leads to more reliable forecasts and efficient inventory management.
**Supplier Relationship Management** : Enhanced communication and negotiation skills, combined with advanced data analytics, enable LLMs to provide valuable insights into supplier performance and predict potential disruptions, strengthening supplier relationships.
**Optimization of logistics and transport** : LLMs analyze routes and predict delays, propose alternative routes, and optimize delivery times and costs. The result is more efficient transport management and reduced operating costs.

### Improve decision-making

LLMs facilitate real-time data analysis and provide predictive insights that improve decision-making processes. They support scenario planning, risk management, and automated decision support systems, enabling supply chain managers to make informed and proactive decisions.

### Customization and customization

LLMs can be adapted to meet the unique challenges of different industries. Adaptive learning capabilities allow these models to continuously integrate new data and improve their performance to ensure that supply chain operations remain resilient and responsive to changing conditions.

## 9- Conclusion

The integration of LLMs into SCM represents a significant leap forward in the industry's ability to optimize operations, reduce costs, and improve decision-making processes. By leveraging the advanced data processing capabilities of LLMs, companies can improve demand forecasting, inventory management, supplier relationship management, and logistics optimization. The synergy between LLMs and emerging technologies such as IoT, blockchain, and robotics is paving the way for the creation of smarter, more autonomous supply chains.

However, the introduction of LLMs in SCM also poses challenges in terms of mitigation bias, data protection, and the need to train staff to manage new AI-driven processes. Taking these ethical considerations into account is essential to ensure fair and transparent AI practices.

To maximize the benefits of LLMs, SCM professionals must invest in high-quality data management, encourage cross-functional collaboration, and align LLM initiatives with overall business goals. In this way, companies can drive innovation, sustainability, and competitive advantage in the ever-changing landscape of supply chain management. The potential of LLMs to revolutionize SCM is immense and offers strategic benefits that can lead to more resilient and responsive supply chains in the face of dynamic market conditions.